\documentclass[preprint]{aastex}
\usepackage{natbib}                
\usepackage{graphicx}              
\usepackage[usenames,dvipsnames,svgnames]{xcolor}
\usepackage{ulem}

\newcommand{\sect}[1]{Section~\ref{sect:#1}}

\newcommand{\fig}[1]{Figure~\ref{fig:#1}}
\newcommand{\figs}[2]{Figures~\ref{fig:#1} and \ref{fig:#2}}

\newcommand{\curl}{ \vec{\nabla} \times}
\newcommand{\degree}{\ensuremath{^\circ}}

\newcommand{\BE}{\begin{equation}}
\newcommand{\EE}{\end{equation}}
\newcommand{\BA}{\begin{eqnarray}}
\newcommand{\EA}{\end{eqnarray}}

\newcommand{\eg}{\textit{e.g.}}

\shorttitle{Quasi-separatrix Layers and Active Region Upflows}
\shortauthors{Mandrini et al.}

\begin{document}

\title{Parallel Evolution of Quasi-separatrix Layers and Active Region Upflows}
\author{C.H. Mandrini\altaffilmark{1,2},  D. Baker\altaffilmark{3}, P. D\'{e}moulin\altaffilmark{4},
G.D. {Cristiani}\altaffilmark{1,2}, L. van Driel-Gesztelyi\altaffilmark{3,4,5},
S. Vargas Dom\'\i nguez\altaffilmark{6}, F.A. Nuevo\altaffilmark{1,2}, A.M. V\'asquez\altaffilmark{1,2} and  M. Pick\altaffilmark{4}}
\altaffiltext{1}{Instituto de Astronom\'\i a y F\'\i sica del Espacio (IAFE), UBA-CONICET, CC. 67, Suc. 28 Buenos Aires, 1428, Argentina}
\altaffiltext{2}{Facultad de Ciencias Exactas y Naturales (FCEN), UBA, Buenos Aires, Argentina}
\altaffiltext{3}{UCL-Mullard Space Science Laboratory, Holmbury St Mary, Dorking, Surrey, RH5 6NT, UK}
\altaffiltext{4}{Observatoire de Paris, LESIA, UMR 8109 (CNRS), F-92195 Meudon Principal Cedex, France}
\altaffiltext{5}{Konkoly Observatory, Research Centre for Astronomy and Earth Sciences, Hungarian Academy
                of Sciences, Budapest, Hungary}
\altaffiltext{6}{Observatorio Astron\'omico Nacional, Universidad Nacional de Colombia, Bogot\'a, Colombia}

\begin{abstract}
Persistent plasma upflows were observed with Hinode's EUV Imaging Spectrometer (EIS) at the edges of active region (AR) 10978 as it crossed the solar disk.
We analyze the evolution of the photospheric magnetic and velocity fields of the AR, model its coronal magnetic field, and compute the location of magnetic null-points and quasi-sepratrix layers (QSLs) searching for the origin of EIS upflows.  Magnetic reconnection at the computed null points cannot explain all of the observed EIS upflow regions. However, EIS upflows and QSLs are found to evolve in parallel,  both temporarily and spatially. Sections of two sets of QSLs, called outer and inner, are found associated to EIS upflow streams having  different characteristics.
The reconnection process in the outer QSLs is forced by a large-scale photospheric flow pattern which is present in the AR for several days. We propose a scenario in which upflows are observed provided a large enough asymmetry in plasma pressure exists between the pre-reconnection loops and for as long as a photospheric forcing is at work. A similar mechanism operates
in the inner QSLs; in this case, it is forced by the emergence and evolution of the bipoles between the two main AR polarities.
Our findings
provide strong support to the results from previous individual case studies investigating the role of magnetic reconnection at QSLs as the origin of the upflowing plasma.
Furthermore, we propose that persistent reconnection along QSLs does not only drive the EIS upflows, but it is also responsible for a continuous metric radio noise-storm observed in AR 10978 along its disk transit by the Nan\c cay Radio Heliograph.
\end{abstract}


\section{Introduction}
\label{sect:Intro}

The EUV Imaging Spectrometer
\citep[EIS:][]{Culhane07}, onboard the Hinode satellite \citep{Kosugi07},  has provided observations of the presence of plasma flows in various solar environments,
from coronal holes to active regions (ARs). One of the most remarkable EIS results was the finding of ubiquitous plasma
upflows seen at the borders of ARs \citep{Harra08}. These are located in regions of low electron density, low radiance, and over monopolar areas
\citep{DelZanna08, Harra08, Doschek08}.  They have been observed
to persist at nearly the same location from a day to at least one week \citep{Doschek08, Demoulin13}.
The Fe {\scshape xii} 195.12 \AA\ blueshifted line-of-sight velocities typically range from a few to 50 km s$^{-1}$ and are faster in hotter coronal emission lines.
\citet{Demoulin13} carried out a detailed analysis of the evolution of upflows in an AR during its disk transit. They concluded that the global temporal variation of the velocities was consistent with a quasi-static flow subjected to a projection effect along the line-of-sight on upflows tilted from the vertical away from the AR core.

Several driving mechanisms have been proposed to explain their origin \citep[see][and references therein]{Baker09a}. In particular,
noticing that in the analyzed examples the upflows appeared at locations where magnetic field lines with drastically different
connectivities were anchored, \citet{Baker09a} proposed that in their case study (AR 10942)
magnetic reconnection between closed field lines of the AR and either large-scale externally connected or ``open'' field lines was a viable
mechanism for driving the upflows.

 \citet{DelZanna11} analyzed EIS upflows in two ARs (AR 10961 and 10955) and found a null point high above in the coronal field in both ARs. These authors suggested that the continuous growth of the ARs maintained a steady reconnection process at the null point. In their view, interchange reconnection occurred between closed, high-density loops in the core of the AR and neighboring open, low-density flux tubes. In this way, magnetic reconnection created a strong pressure imbalance which was the main driver of the plasma upflows \citep{Bradshaw11}.

As in flares, upflows are not only related to reconnection at null points but also at quasi-separatrix layers (QSLs) \citep[see the examples studied by][]{Baker09a,vanDriel-Gesztelyi12}. QSLs are defined as thin volumes in which field lines display strong connectivity
gradients \citep[][]{Demoulin96}. When these gradients become infinitely large, a QSL becomes a separatrix.
QSLs, like separatrices, are places where strong currents can form during the evolution of a magnetic field having a high Lundquist number \citep{Aulanier05,Buchner06,Effenberger11,Janvier14b}.  Therefore, QSLs are natural locations where magnetic reconnection can take place \citep[][]{Priest95, Demoulin96}.  This was confirmed by MHD numerical simulations \citep{Milano99,Aulanier06,Aulanier10,Wilmot-Smith10,Janvier13}, by kinetic numerical simulations \citep{Wendel13,Finn14}, and in laboratory plasmas \citep{Lawrence09,Gekelman12}.   \citet{Hesse88} and \citet{Schindler88} developed a general framework for three-dimensional (3D) reconnection based on the description of the magnetic field using Euler potentials and localized non-ideal regions. QSLs are the locations where these non-idealnesses can occur; therefore, the two approaches are complementary \citep{Demoulin96b,Richardson12}.

 Separatrices and null points are also embedded in QSLs in which reconnection complies with special properties, such as the slippage of field lines \citep{Masson09,Masson12}.  Complex magnetic configurations involving both a null point and QSLs have been found to be associated with some observed upflows.
In a quadrupolar configuration, formed by
AR 10980 and a neighbouring magnetic bipole,
\citet{vanDriel-Gesztelyi12} showed
that plasma upflows observed with EIS were co-spatial with QSL locations, including the separatrix of a null point for a fraction of the upflows. Global potential-field source-surface (PFSS) modeling indicated that part of the upflowing EIS plasma could access the solar wind along reconnected field lines, which extended up to the source surface, passing through the vicinity of the null point.

Upflowing plasma from other ARs may not have direct access to the solar wind, however. For example, AR 10978 was
an isolated bipolar region that, according to a PFSS model, was completely covered by the separatrix surface of a helmet streamer \citep{Culhane14}. \citet{Brooks11,Brooks12} analyzed EIS upflows and
found that the abundance of Si was always enhanced over that of S by a factor of 3\,--\,4 (a classical value for FIP-bias enhancement in the corona).
When the AR's western side was oriented in the Earth direction, the Si/S ratio, measured with the Solar Wind Ion Composition Spectrometer \citep[SWICS,][]{Gloeckler98} onboard the Advanced Composition Explorer (ACE) a few days later, was found comparable to the Si/S abundance ratio measured in the corona.  This provided evidence to support a connection between the solar wind and the coronal plasma in the upflow region.
 \citet{Culhane14} concluded that, even though AR 10978 was isolated and completely covered by closed streamer field-lines, the coherent magnetic field, proton velocity, and density variation at L1, together with the matching FIP-bias evolution at the Sun and L1, were clear proof of the presence of AR plasma in the slow solar wind. Based on a global topology computation and analysis of noise-storm radio signatures, \citet{Mandrini14a} proposed that the AR plasma could reach the solar wind via a two-step reconnection process.
The first step was proposed to occur in the AR between closed AR loops and long externally connected loops, while the second step was shown to involve the large-scale global coronal field at a high altitude null point. Only this second step was analyzed in depth by \citet{Mandrini14a}, while the present work completes the previous study focussing on the role of QSLs during the first reconnection step.

The spatial relation between upflow and QSL locations and magnetic field-line traces and connectivity, led \citet{Baker09a} and \citet{vanDriel-Gesztelyi12} to suggest that magnetic reconnection at QSLs was at the origin of EIS upflowing plasma. In this article we put forward a proof of the concept. We analyze the temporal evolution of EIS upflows in relation to QSL evolution as AR 10978 crosses the solar disk during Carrington rotation (CR) 2064. Our results indicate that the evolution of the AR magnetic field leads to the evolution of QSL locations, which in turn leads to a spatial evolution of EIS upflows.

Our article is organized as follows. In \sect{magnetic-evol} we describe the evolution of the photospheric magnetic and velocity fields of AR 10978. We briefly discuss the EIS upflow evolution in \sect{upflow-evol}.
  In \sect{model-topo}, we model the AR coronal field (\sect{model}) and search for the presence of magnetic null
points (\sect{nulls}).
  Since reconnection at nulls cannot explain the observed upflows, we further analyze QSLs in \sect{qsls}. In particular, we demonstrate the spatial and temporal relation between upflow regions and QSLs (\sect{qsls-evol}).
Based on this analysis, combined with the photospheric magnetic field evolution, we unveil the characteristics of upflows originating from different AR locations (\sect{qsls-detail}).
  Next, in \sect{radio}, we find the presence of weak noise storms that remain located above the AR during its whole disk transit. We associate their origin to the magnetic reconnection process occurring at the AR QSLs.
  Finally, in \sect{conclusions}, we summarize our results and draw our conclusions.


\begin{figure}  
\centerline{\includegraphics[width=0.5\textwidth]{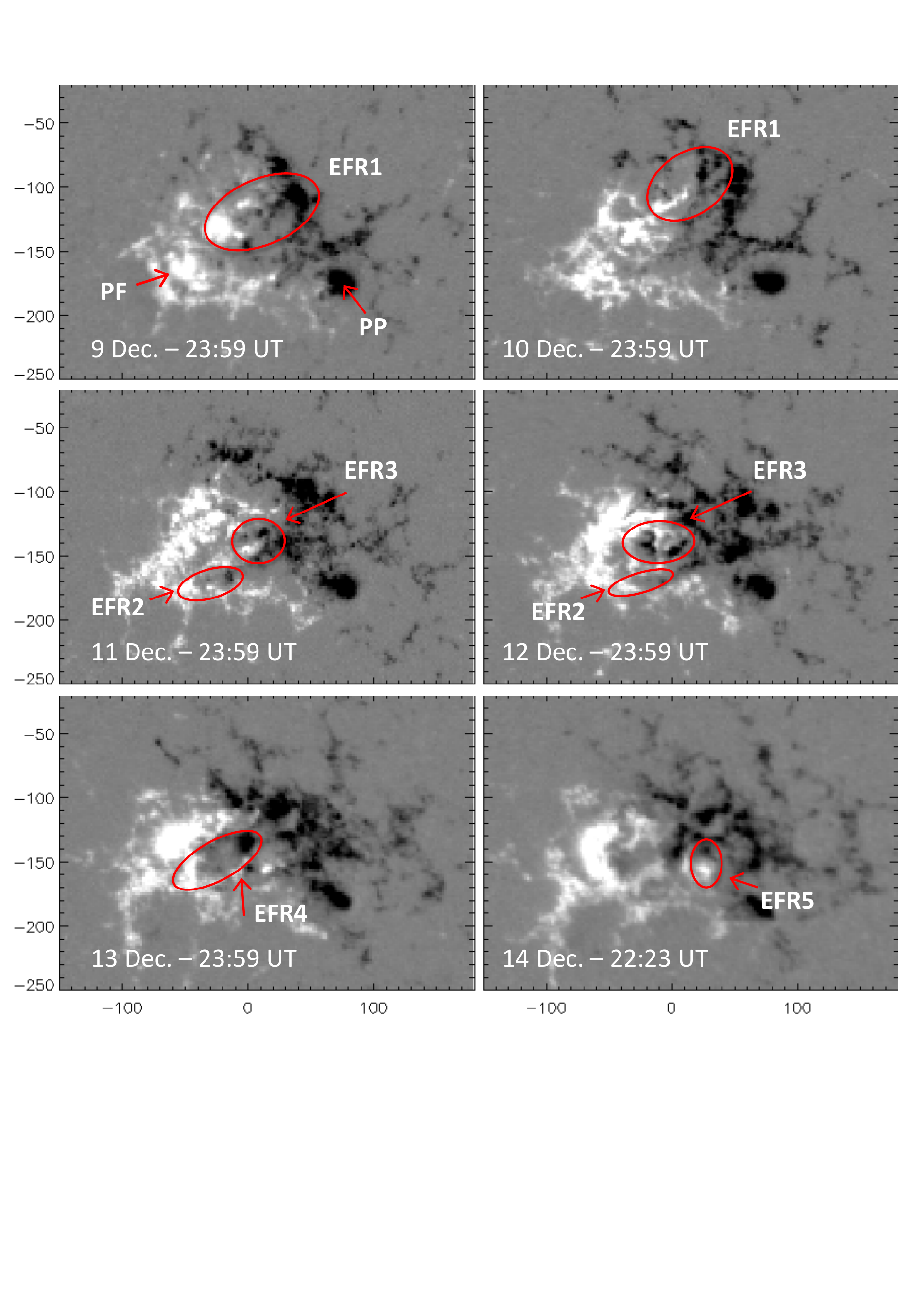}}
\vspace*{-2.cm}
\caption{Magnetic field evolution of AR 10978 for six days from 9 to 14 December 2007. Five flux emergence episodes, indicated as EFR1\,--\,EFR5, are observed along this period. They occurred between the two pre-existing polarities marked as PF and PP, which emerged on the far side of the Sun. The saturation level is 800 G and the axes are in arcsec. The vertical axis is the AR location in the solar north-south direction with the origin corresponding to the solar equator, while the horizontal axis indicates the east-west coordinates derotated to the time when the AR was close to the central meridian.
}
\label{fig:mdi-evol}
\end{figure}  

\begin{figure*}  
\centerline{\includegraphics[width=0.95\textwidth]{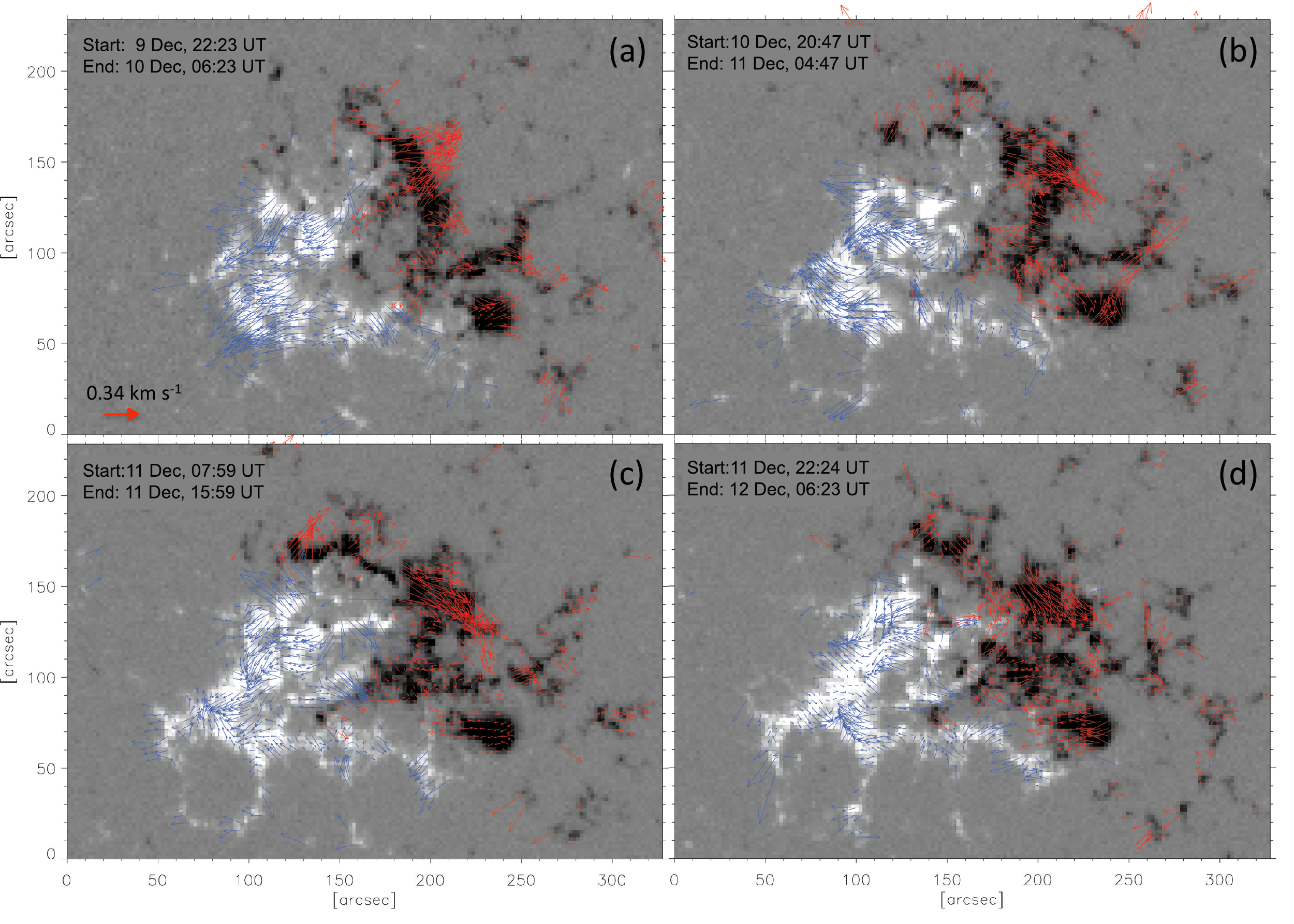}}
\caption{Maps of the photospheric transverse velocities displayed over magnetograms.
The flow maps, generated from a sequence of magnetograms previously derotated to central meridian passage position, are derived from an LCT analysis in the interval indicated in each map (top-left corner) employing a FWHM correlation window of 10\arcsec . Red and blue arrows represent velocities of up to 0.5 km s$^{-1}$, the segment at the bottom left corner of panel (a) indicates
the velocity scale.  These velocities are averaged over the time intervals indicated in each panel. Notice that the velocity vectors show a global separation of the leading and following polarities as expected in a mature AR. The magnetic field background corresponds to the average image computed in the flow-map time interval. Positive (white) and negative (black) polarities are saturated at $\pm$ 800 G. The field of view is the same as in \fig{mdi-evol} with the origin set at the lower left corner.}
\label{fig:flows-evol}
\end{figure*}  

\section{Evolution of AR 10978 during its Disk Transit}
\label{sect:evolution}

\subsection{Magnetic Field Evolution}
\label{sect:magnetic-evol}

Active region 10978, observed with the Michelson Doppler Imager \citep[MDI,][]{Scherrer95} onboard the {\it Solar and Heliospheric Observatory} (SOHO), rotated onto the disk on 7 December 2007. By this time, it was a mature AR. From its state of evolution and the separation of its opposite-sign polarity spots, it is likely that the AR was at least 3\,--\,5 days old when it appeared on the east solar limb \citep[][]{vanDriel15}.

We checked data from the {\it Solar-Terrestrial Relations Observatory} spacecraft (STEREO-A and B) in order to constrain its emergence time using data from the Extreme-ultraviolet Imager \citep[EUVI,][]{Wuelser04}. On the date of the AR appearance, STEREO A and B were separated from the Sun-Earth line by about 21$\degree$ in each direction. In the 195 \AA\ data of STEREO-B, the AR coronal loops could be well seen over the east limb already on 5 December. STEREO-A data showed a potential first sign of flux emergence at the future location of AR 10978 on 21\,--\,22 November at the west limb, indicated by a significant increase of coronal emission from that location (\url{http://stereo-ssc.nascom.nasa.gov/cgi-bin/images}). Therefore, the first indications of the AR appearance could have been as old as 18 days when its magnetic field was first mapped with MDI.
However, this early flux emergence (on 21\,--\,22 November) might not have been a major episode and, perhaps, the flux decayed quickly.

\begin{figure*}  
\centerline{\includegraphics[width=1.\textwidth]{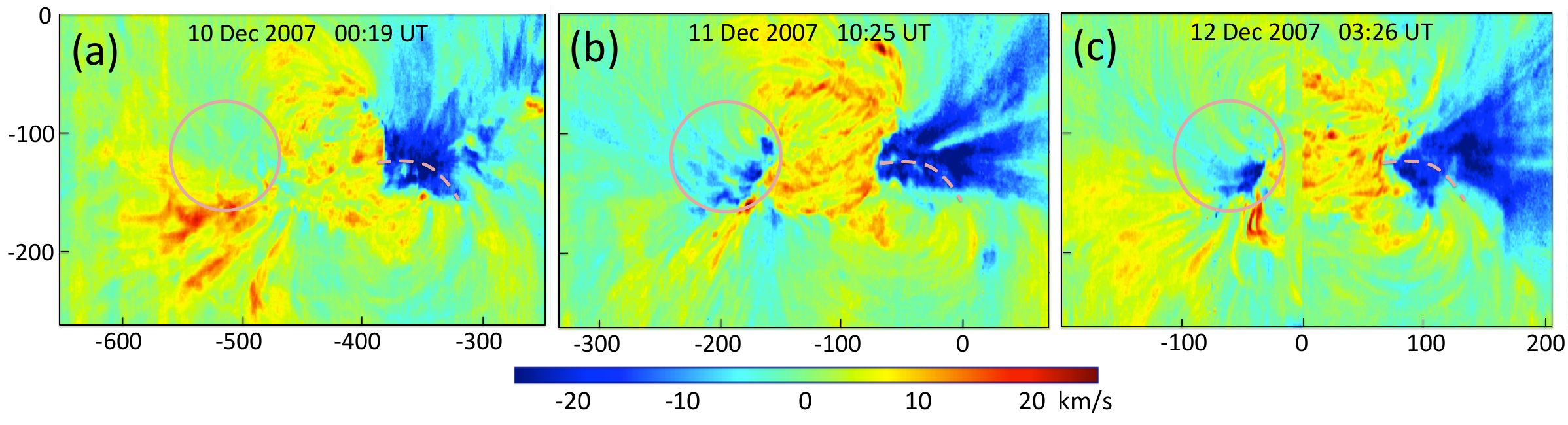}}
\caption{Doppler velocity maps around AR 10978 CMP obtained from the Fe {\scshape xii} 195.12~\AA\ EIS emission-line profiles. The color bar at the bottom indicates the velocity scale. The full pink circle surrounds low velocity regions, while the dashed pink lines separate upflow streams with different characteristics. Additional Doppler maps obtained with EIS have been published by \citet{Brooks11} and \citet{Demoulin13}. The vertical and horizontal axes indicate the position on the Sun in arcsec.}
\label{fig:upflow-evol}
\end{figure*}  

AR 10978 had a peak mean magnetic flux of $3 \times 10^{22}$ Mx \citep{Mandrini14a}, which corresponds to the large AR category \citep[see e.g.][]{vanDriel15}. Such ARs may survive several months; indeed, AR 10978 returned in the following rotation as AR 10980 \citep[see e.g.][]{vanDriel-Gesztelyi12} and its location was magnetically active for several more solar rotations.

From 9 to 13 December, magnetic flux emergence continued within the AR (\fig{mdi-evol}). We were able to follow five significant flux emergence episodes, which are indicated in the figure with red ellipses and numbered as EFR1\,--\,EFR5 (EFR, emerging flux region). These occurred between the pre-existing polarities marked as PP and PF in the top left panel of \fig{mdi-evol}.
Emerging bipoles are characterized by the divergence of opposite-sign polarity flux concentrations; therefore, these EFR sites are locations of fast photospheric motions. The evolution of the border of the supergranule to the north of PP is also noteworthy. This supergranule evolved as minor polarities emerged within it (not visible in \fig{mdi-evol} because of the chosen high saturation level). By 11 December, the supergranule border is no longer visible.

We have computed the transverse flows of the photospheric magnetic field features employing a local correlation tracking technique \citep[LCT,][]{November88}. The proper motions of the magnetic elements over the MDI sequence of magnetograms (spatial resolution 1.98\arcsec ) was computed using a Gaussian tracking window with full width at half maximum (FWHM) of 10\arcsec .

Different values for the correlation window were used, varying from 6 to 14\arcsec . The general pattern of the horizontal velocity vectors did not vary significantly from case to case, in the sense that we obtained similar results in terms of their distribution by direct visual inspection of the computed flow maps. Therefore, the size of the correlation window (10\arcsec ) was selected following a criterion based on reducing the level of noise and producing a coherent tracking of the magnetic features. Such window size is within the interval of sizes used in previous works \citep[\eg ][]{Nindos03,Chae04,Vemareddy12}.

The time cadence of the LCT, 96 min, is fixed by the observations. It
filters the evolution of faster time scales and, in particular, the amplitude of the deduced velocity decreases \citep[\eg\ see Figure 7 of ][]{Chae04}. However, since the upflows we analyze evolve on a time scale of days, we expect the 96 min cadence to be sufficient to derive the photospheric flows relevant to understand these upflows. The time series for the analysis spans from 22:23 UT on 9 December 2007 to 06:23 UT on 11 December 2007.
Prior to applying the LCT method, the sequence of images was aligned to eliminate the possible jitter and the solar rotation of the observed target within the field of view (FOV), which results in flow maps derotated to the AR central meridian passage position. Finally, it should be noted that whereas the LCT results give a reliable characterization of the flow patterns, the absolute values of the transverse velocities should be taken with caution because this technique, in general, underestimates them \citep{November88,Vargas2008}.

The maps of the transverse flows in \fig{flows-evol} display the velocities, using arrows, for pixels with magnetic field values over a threshold of 800~G. The computed flows show a global diverging pattern of the main photospheric magnetic polarities.
Emergence episodes are also detected in the flow map, see e.g. the one corresponding to EFR1 in \fig{flows-evol}a. Another noticeable flow pattern is that on the western portion of the northern negative polarity (\fig{flows-evol}b-d). It indicates a motion of the magnetic features from the north-east to the south-west.

\subsection{EUV Upflow Evolution}
\label{sect:upflow-evol}

EIS is a raster-scanning instrument capable of constructing large fields of view, up to $\approx$ 600$\arcsec$ in the dispersion direction and 512$\arcsec$ in the slit direction, using the 1$\arcsec$ and 2$\arcsec$ slits and the 40$\arcsec$ slot.  It has a spectral resolution of 22.3 m\AA\ and 1$\arcsec$ pixels. EIS observes in two wavelength ranges 170\,--\,210 \AA\ and 250\,--\,290 \AA.

\begin{figure*} 
\centerline{\includegraphics[width=0.90\textwidth]{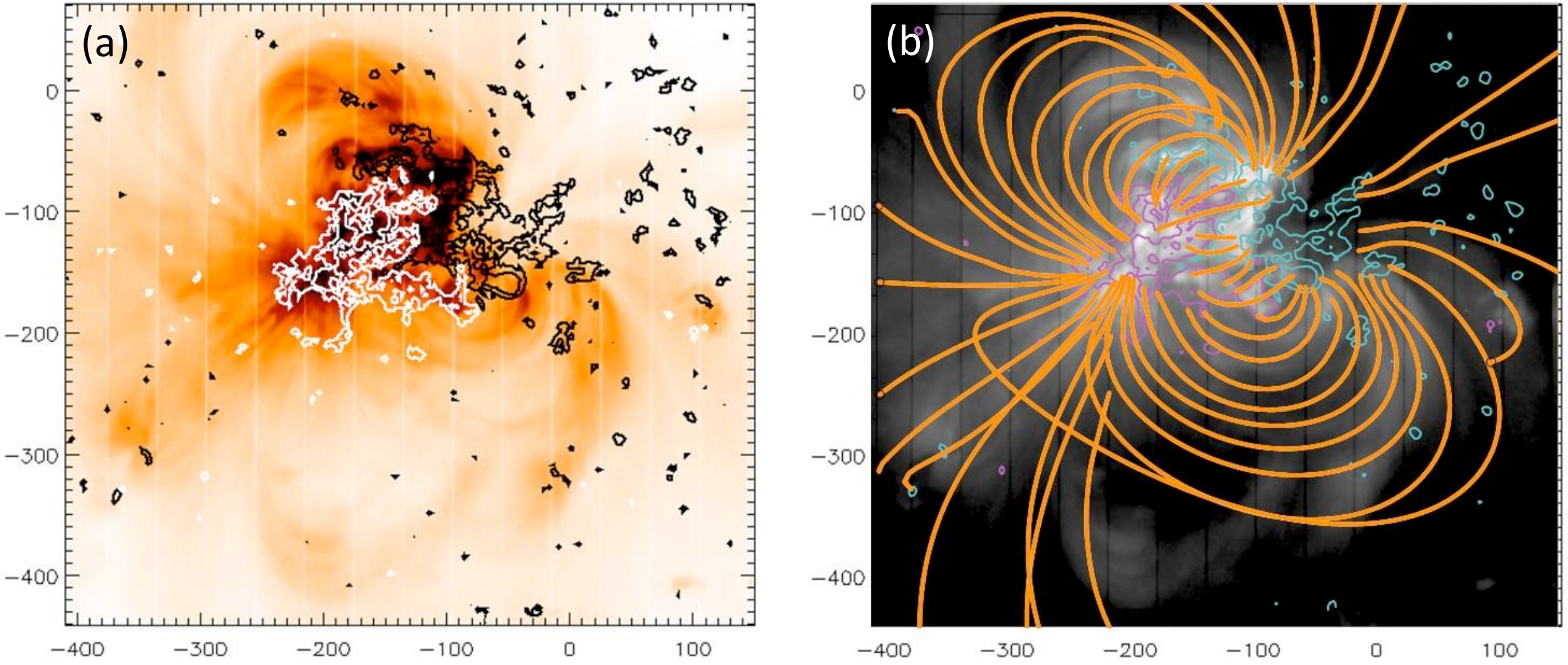}}
\caption{(a) EIS Fe {\scshape xii} emission line intensity map of AR 10978 on 11 December at 10:25 UT overlaid with MDI contours with the same values as those in the right panel (drawn in white (black) for positive (negative) values).
 (b) Modeled MDI magnetic field contours ($\pm$ 100, 500 G, positive (negative) shown in magenta (blue) color) on 11 December at 11:11 UT overlaid on the EIS slot image shown in gray scale. A set of computed field lines matching the global shape of the observed coronal loops has been added in continuous line and orange color.
The vertical and horizontal axes indicate the position on the Sun in arcsec. }
\label{fig:model}
\end{figure*}

AR 10978 is selected for this study because, with the exception of a few other ARs \citep[see e.g.,][]{DelZanna08,DelZanna11}, it has the best EIS spatial and temporal coverage of an AR from limb to limb. This AR was tracked from 6 to 19 December 2007 using the full complement of EIS slits. In this article, we use a set of five large FOV slit rasters around the AR central meridian passage (CMP, see \figs{upflow-evol}{qsls-evol}), which occurred on 11 December at approximately 22:00 UT. The fields of view cover both of the main AR polarities (see Table 1 in \citet{Demoulin13} for more details).
We selected EIS observations of the Fe {\scshape xii} emission line at 195.12 \AA\ ($T \approx$ 1.4 MK) because it is the strongest line within EIS's wavelength ranges. We have also used EIS 40$\arcsec$ slot rasters of the AR that allow us to observe the bright coronal loops, which extend beyond the fields of view of the slit rasters and are useful to constrain the free parameter of our magnetic field model (see \fig{model}).

EIS slit and slot raster data were processed using standard SolarSoft EIS routines to correct
for dark current, cosmic rays, hot, warm and dusty pixels and to remove instrumental
effects of slit tilt and orbital variation in the line centroid position due to thermal
drift.  Doppler velocities for slit rasters were calculated by fitting a single-Gaussian function to the calibrated Fe {\scshape xii} spectra in order to obtain the line center for each spectral profile.  Reference wavelengths
were determined using the average wavelength value of a relatively quiescent Sun patch within each raster.  Velocity maps follow the standard convention of blueshifts (redshifts) corresponding to negative (positive)
Doppler velocity shifts along the line of sight (see \figs{upflow-evol}{qsls-evol}).

As an example, we show in \fig{upflow-evol} three Doppler velocity maps close to the AR's CMP. These are drawn using standard IDL routines. Following \citet{Demoulin13}, we have added guide marks in \fig{upflow-evol} that help
to visualize and track the main upflow structures. The pink circles surround weak velocity patterns (because of projection effect), while the pink dashed lines separate flow streams.

When looking at the upflow pattern in \fig{upflow-evol}, a clear evolution is evident. This was interpreted by \citet{Demoulin13} to be the signature of a projection effect on steady upflows
that were inclined from the vertical at an angle that was larger to the east than to the west \citep[see Figure~14 in][]{Demoulin13}.

\section{Magnetic Field Configuration}
\label{sect:model-topo}

\subsection{Magnetic Field Model of AR 10978}
\label{sect:model}

To compute the magnetic field topology of an AR, we first model its coronal field. We extrapolate the line-of-sight magnetic field
of AR 10978 to the corona using the discrete fast Fourier transform method \citep[as discussed by][]{Alissandrakis81} under the linear force-free field (LFFF) hypothesis ($\curl \vec B = \alpha \vec B$, with $\alpha$ constant).
An example of extrapolation is shown in \fig{model}b
when the AR is at disk center on 11 December 2007. We use as the boundary condition for the magnetic model, the MDI magnetogram at 11:11~UT. This magnetic map is the closest in time to the EIS slot image in Fe {\scshape xii} 195.12~\AA\ (\fig{model}a), which is large enough to identify the global shape of the coronal loops. The value of the free parameter of the model, $\alpha$, is set to best match the observed loops following the procedure discussed by \citet{Green02}. The best-matching
value is $\alpha$ = -3.1 $\times$ 10$^{-3}$ Mm$^{-1}$ (\fig{model}b).

A region four times larger than that encompassed by AR 10978, and centered on it, is selected from each MDI full-disk magnetogram as the boundary condition for each extrapolation.
This magnetic map is embedded within a region twice larger padded with a null vertical field component for two reasons. First, to decrease the modification of the magnetic field values since the method to model the coronal field imposes flux balance on the full photospheric boundary (i.e., the flux unbalance is uniformly spread on a larger area, so the removed uniform field is weaker as the area is larger).
Second, to decrease aliasing effects resulting from the periodic boundary conditions used on the lateral boundaries of the coronal volume. The photospheric boundary condition is then written in a $1024 \times 1024$ horizontal grid
to maintain the spatial resolution of the observations. This lets us distinguish field lines that connect to the surrounding quiet-Sun regions from those that are potentially `open' lines as they leave the extrapolation box.

To compute the topology at different times during the AR transit, we use the same value of $\alpha$ with the corresponding MDI map as
the boundary condition.  We have checked that the large scale loops observed in EIS slot images on the dates used in our analysis can be globally fitted with this $\alpha$ value. Indeed, the magnetic configuration of AR 10978 has a low shear and the coronal magnetic field configuration is mainly evolving due to the evolution of the photospheric boundary condition rather than to the change in the $\alpha$ value. We also carry out a transformation of coordinates from the local AR frame to the observed one \citep[see][]{Demoulin97} to obtain a model for which the QSL locations can be compared to EIS upflows (\sect{qsls}).

\subsection{Magnetic Null Points in the AR Neighborhood}
\label{sect:nulls}

The origin of some EIS upflows has been attributed, either directly or indirectly, to magnetic reconnection
occurring in the vicinity of coronal magnetic null points \citep{DelZanna11,vanDriel-Gesztelyi12,Mandrini14a}. Therefore, after summarizing the main properties of null points in the next paragraph, we investigate if the observed upflows in AR 10978 are due to magnetic reconnection at these null points.

The field connectivity around a null point is
characterized by the presence of so-called spines and fans \citep[see e.g.,][]{Longcope05,Pontin11}.  A fan surface separates the coronal volume into two connectivity domains, while all field lines in the close vicinity of the fan converge to the associated spine line.
The local field connectivity around a null point can be found using the linear term of the Taylor expansion of the magnetic field \citep[see][and references therein]{Demoulin94b,Mandrini06}. From the diagonalization of the Jacobian matrix of the field, one finds three eigenvectors and the corresponding eigenvalues, which add up to zero to locally satisfy the field divergence-free condition. The eigenvalues are real for coronal conditions \citep{Lau90}. A positive null point has two positive fan eigenvalues and conversely for a negative null. The fan surface is defined by all of the field lines starting at an infinitesimal distance from the null in the plane defined by the two eigenvectors that correspond to the eigenvalues that have the same sign. In a similar way, the spines are defined by the field lines tangent to the third eigenvector.

\begin{figure} 
\centerline{\includegraphics[width=0.5\textwidth]{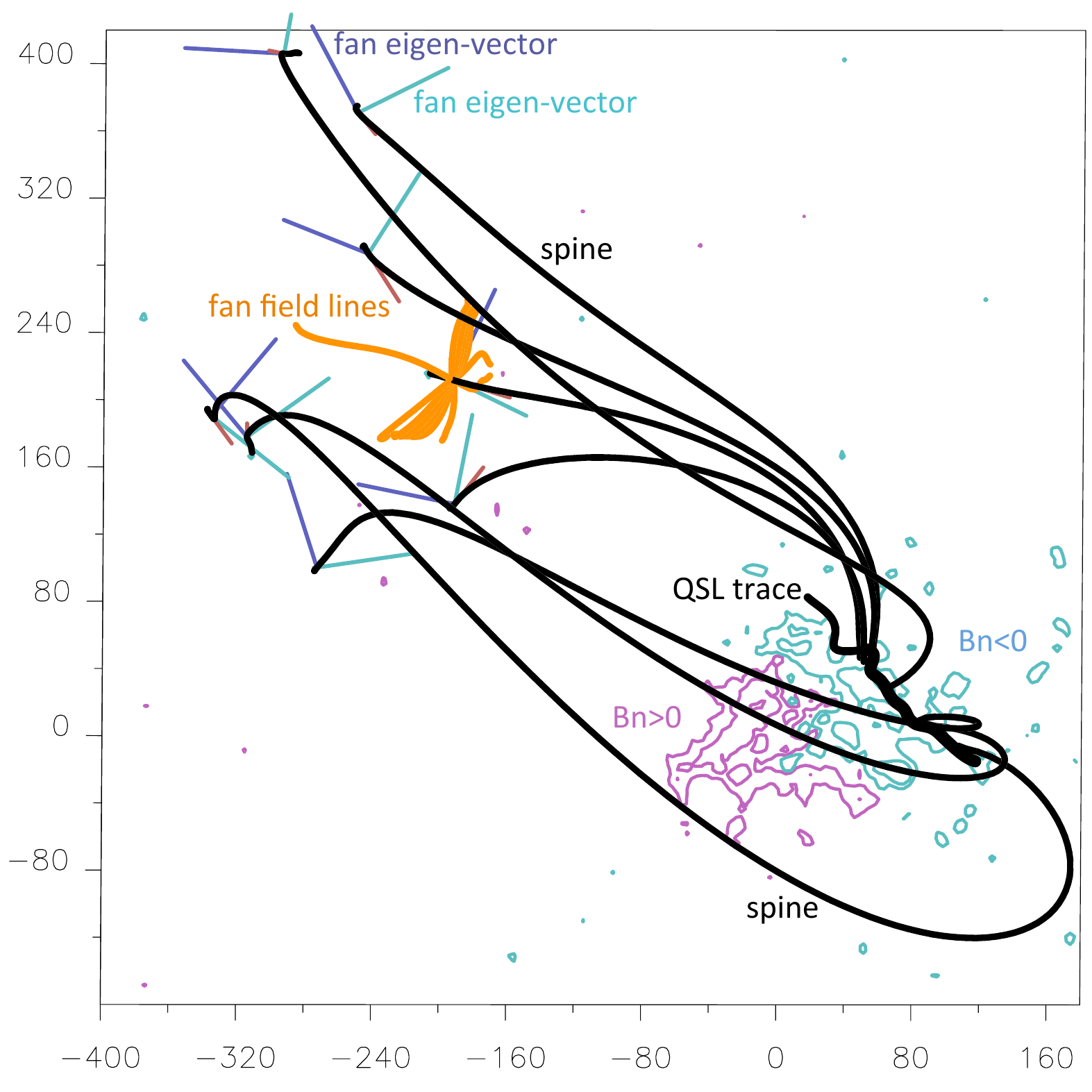}}
\caption{Magnetic field model of AR 10978 and null points with a topological structure linked to its main polarities. The model corresponds to the MDI magnetic map at 03:11 UT on 12 December 2007. The locations of all magnetic
null-points at heights greater than $\approx$ 20 Mm above the photosphere are shown as the intersection
of three colored segments (see text). We have drawn the null-point spines as black continuous lines and have added a set of field lines (orange continuous lines) in the vicinity of the fan plane below one of the nulls. The external western QSL trace, depicted as a thick black continuous line, overlays the $B_n < 0$ polarity  (see \sect{qsls} and its text). The axes in this panel are in Mm, with the origin set in the AR and
the isocontours of the line-of-sight field correspond to $\pm$100, $\pm$500 G in continuous magenta (blue) style for the positive
(negative) values.}
\label{fig:nulls}
\end{figure}

We computed the location of magnetic null points for the MDI magnetograms closest in time to the EIS velocity maps described in \sect{upflow-evol}.  None of the null points found were located above the AR and most of the ones above quiet-Sun regions had no field lines connecting to the AR polarities. \fig{nulls} illustrates the location of all magnetic null points at heights above 20 Mm, which are related to AR 10978 for an MDI magnetogram on 12 December. The location of the null points is indicated by the intersection of three segments that correspond to the direction of the three eigenvectors of the Jacobian matrix. These segments are color coded to indicate the magnitude of the corresponding eigenvalue.
For a negative null, dark blue (light blue) corresponds to the highest
(lowest) negative eigenvalue in the fan plane and red to the spine eigenvalue. All null points in \fig{nulls} are negative.
Their spines, drawn in black in the figure, are connected to a quiet-Sun
polarity at one end and to the main negative polarity of AR 10978 at the other.  As an example, we have drawn in orange field lines belonging to the fan of one of the null points.
Since these nulls are above the quiet Sun, they are associated to a magnetic field intensity, $B$, lower than in the AR ($B$ lies in the range $[20,50]$~G). Such field values are present in small polarities with sizes $\approx 6$ pixels, so they are not within the magnetogram noise. In fact, the height of these nulls, above 20 Mm, confirms that they are associated to real local magnetic polarities.

\begin{figure*} 
\centerline{\includegraphics[width=0.75\textwidth]{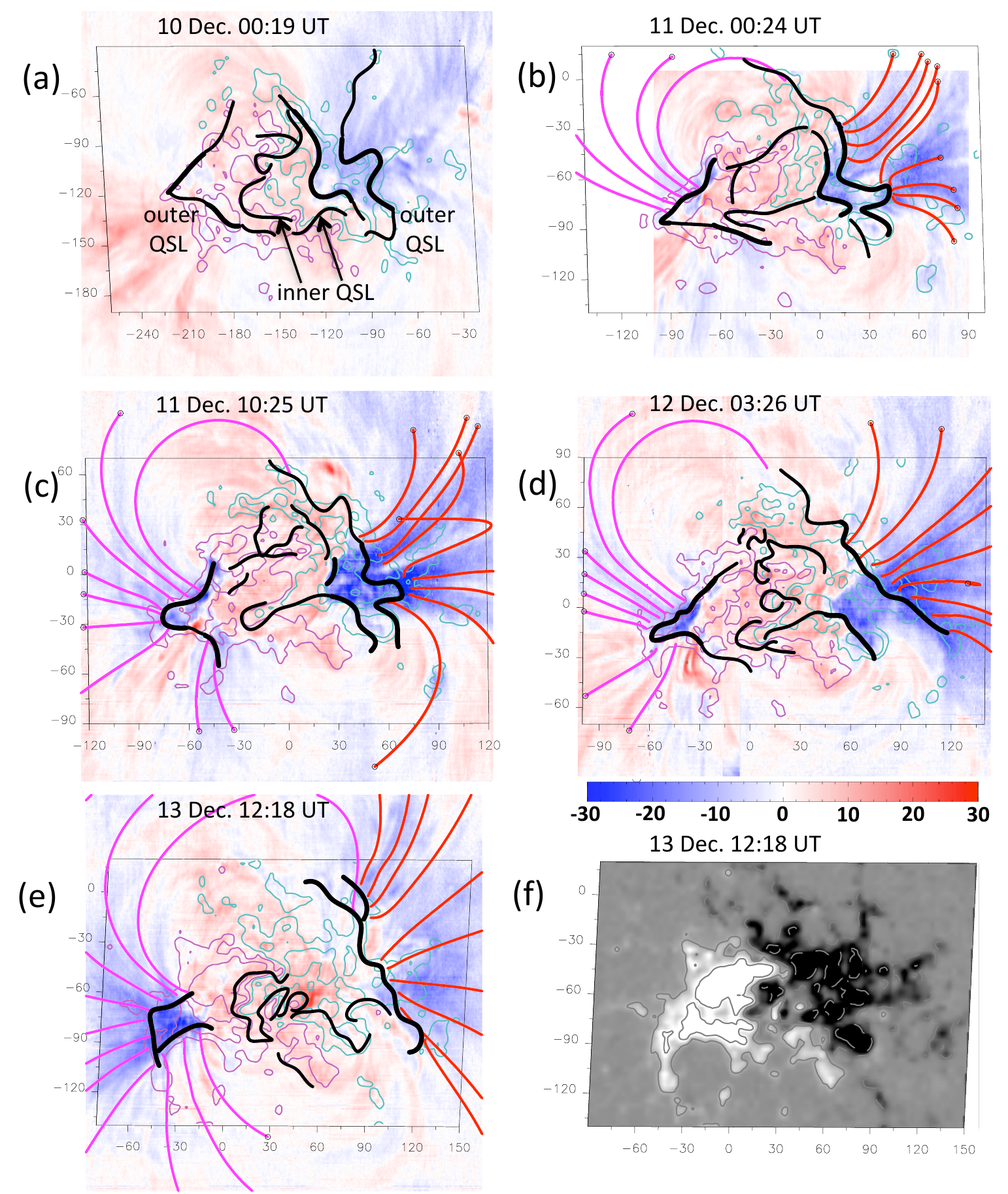}}
\caption{Evolution of the topology of AR 10978 for a set of magnetograms from 10 to 13 December 2007 (panels (a) to (e)). QSL traces related to the EIS upflows are marked with thicker black lines, while those not related are indicated with thinner ones. Most of these thinner black traces mark the drastic connectivity changes due to the presence of minor emerging bipoles in between the main AR positive and negative polarities. In panel (a) we have labeled the traces that we call outer and inner QSLs.
The upflow (downflow) regions, for the Fe {\scshape xii} EIS emission line, are indicated as blue (red) shaded zones. The color bar below panel (d) indicates the velocity scale in km s$^{-1}$. We have also added sets of computed field lines anchored in the vicinity of the outer QSL traces in panels (b) to (e) (see \sect{qsls-evol} and \fig{reconnection} for their color convention). Field lines ending in a circle reach the borders of the box selected for the drawing. The convention for the magnetic field isocontours, in panels (a) to (e), and panel axes are the same as in \fig{nulls}.
Panel (f) depicts the line-of-sight magnetic field in gray scale with white (black) color for the positive (negative) field values. Two isocontours, with the same values as those in the other panels, have been added in gray continuous (dashed) line for the positive (negative) field. This panel has been added to better visualize the photospheric field distribution. The dates and times at the top of each panel correspond to those of EIS velocity maps.  The axes are in Mm with the origin set in the AR.}
\label{fig:qsls-evol}
\end{figure*}

It is highly improbable that magnetic reconnection at these null points can drive all of the EIS upflows in AR 10978.
On one hand, none of the null points has field lines linked to the upflow region on the eastern AR border. On the other hand, reconnection at these nulls could possibly result in upflows only in the neighborhood of the spines; then, this reconnection process can, at most, explain a fraction of the observed upflows (compare \fig{nulls} to \fig{qsls-evol}d using the QSL trace as a guide).
Furthermore, reconnection at these nulls could only slightly affect the coronal loops since they are embedded in weak magnetic fields, which  implies a low amount of energy release, and the structure of the pre- and post-reconnection loops is almost the same (except in the neighborhood of the nulls). Therefore, reconnection at these nulls, far away from the AR, is not expected to drive the upflows observed at the AR border.

\section{Quasi-Separatrix Layers and upflows}
\label{sect:qsls}

\subsection{QSL characteristics}
\label{sect:qsls-general}

Based on our previous results \citep{Baker09a,vanDriel-Gesztelyi12}, we compute QSLs for a series of magnetic maps to confirm or refute the relationship between EIS upflow and QSLs during the AR evolution.

The method to compute QSLs was first described by \citet{Demoulin96}.
QSLs were defined using the norm, $N$, of the Jacobian matrix of the field-line mapping. This norm depends on the direction selected to compute the mapping; then, $N$ has, in general, different values at both photospheric footpoints of a field line.
To overcome this problem \citet{Titov02} proposed  to introduce a function called the squashing degree, $Q$, which is defined as $N$ to the second power divided by the ratio of the vertical component of the photospheric field at the two opposite field-line footpoints.
$Q$ takes into account only the distortion of the field-line mapping, independently of the field strength, and is invariant along each field line.

To determine the QSL locations we have to integrate a huge number of field lines. A key point is to use a very precise integration method since derivatives of the mapping are needed to calculate $N$ and $Q$. In order to decrease the computation time we use an adaptive mesh, i.e., the mesh is refined iteratively only around the locations where the largest values of $Q$ were found in the previous iteration. The fraction of points retained at each iteration controls the computation speed and how much the finally calculated $Q$-map will extend towards the lower values of $Q$. The iteration at a location is ended when the QSL is locally well resolved or, ultimately, when the limit of the integration precision is reached.  Such computations can be performed at the photospheric level and also within the full coronal volume \citep{Pariat12}.

In complex magnetic configurations (e.g. quadrupolar or multipolar) the location of QSLs is strongly determined by the distribution of the magnetic field polarities at the photosphere
\citep[see][and references therein]{Mandrini14b} and the maximum value of $Q$ is typically very high (many orders of magnitude, up to infinity). As the magnetic field configuration is more bipolar, the QSL locations become more influenced by the presence of magnetic shear and/or twist and the value of $Q$ tends to be lower \citep[see the discussion in][]{Demoulin97}. AR 10978 is an isolated globally bipolar region within which several smaller bipoles emerged during its transit across the disk (see Figure 1 in \citet{Mandrini14a} and \fig{mdi-evol} in this article); this increases the complexity of the QSL pattern and the $Q$ values between the two main polarities.

\subsection{QSL evolution}
\label{sect:qsls-evol}

QSLs are the expected locations where magnetic reconnection can occur efficiently at coronal heights, releasing the stored magnetic energy (see references in \sect{Intro}). Then, the energy released is transported along field lines toward the chromosphere. Indeed, several studies have found that flare brightenings are located along chromospheric QSL traces \citep{Demoulin07,Mandrini10,Aulanier11,Sun13,Dalmasse14,Vemareddy14,Savcheva15}.

In the case of EIS upflows, the relationship between QSLs and upflow regions is more indirect because the upflows are observed over a broad range of coronal heights. A direct comparison between upflow regions and QSL locations, considering also the trace of reconnected field lines, is also complicated because
the EIS velocity maps result from the integration of optically thin emission over a large depth along the line-of sight; this integration is also affected by projection effects.
However, taking into account the results from previous studies \citep{Baker09a,vanDriel-Gesztelyi12}, upflows are expected in the vicinity of QSLs where
short and high-density loops can reconnect with large-scale low-density closed or `open' loops.
After reconnection, the higher pressure plasma from the initial short loops will be injected into newly formed large-scale loops creating a pressure gradient that drives the upflows \citep{Bradshaw11}. This process is clearly illustrated in Figure 5 of \citet{vanDriel-Gesztelyi12}.

\fig{qsls-evol} shows the location of QSLs for a series of MDI magnetograms around AR 10978 CMP and illustrates all the minor bipole emergences (compare to \fig{mdi-evol}). These magnetic maps are also the closest in time to the respective EIS velocity maps in Fe {\scshape xii} obtained during the AR disk transit. Extreme care was taken to coalign EIS data with MDI magnetograms.  This was done not only using the image-header informations but also comparing the position of all structures (e.g., loop traces) with the location of the magnetic polarities.
The QSL traces, black continuous thick and thin lines, have been overlaid on the photospheric magnetograms shown as isocontours of the field and the velocity maps indicated by blue (red) shaded regions corresponding to EIS upflows (downflows).
The value of $Q$ is above 400 for all the traces shown; of course, at QSL locations where the null-point spines are anchored (see \fig{nulls}) the value of $Q$ is extremely high ($Q \ge 10^{12}$).

\fig{qsls-evol} shows that the upflow regions are consistently located in the vicinity of QSLs drawn with thicker black lines on both AR main polarities. In panels (a) through (d), two main QSL traces, called outer and inner, extend along the negative western polarity. Sections of these traces (those drawn with thicker lines) are linked to the upflows.
It is striking how well the inner QSL traces match the projected shape of the inner border of western upflow regions. Furthermore,
the outer QSL shapes match the locations where the upflow velocities change in magnitude \citep[see][and \fig{upflow-evol}]{Demoulin13}, indicating upflow regions with different characteristics.
As the AR and its upflows evolve, see \fig{qsls-evol}e, only the westernmost upflows associated to the outer QSL trace remain.  On the eastern AR border, the outer QSL traces
also match quite well the projected shape of the upflow border. Next,
as new minor bipoles emerge and evolve the shapes of the inner QSLs evolve as well and their complexity increases.

We have also added sets of field lines to all panels in \fig{qsls-evol}, but panel (a) to avoid overloading it. These field lines have been computed starting integration in the vicinity of the outer QSLs, to the west (east) of the one on the main negative (positive) AR polarity. The  projected shape of these field lines matches the spatial extension of the upflows at the borders of the AR providing evidence of the close relationship between upflows and QSLs.

\begin{figure*} 
\centerline{\includegraphics[width=0.90\textwidth]{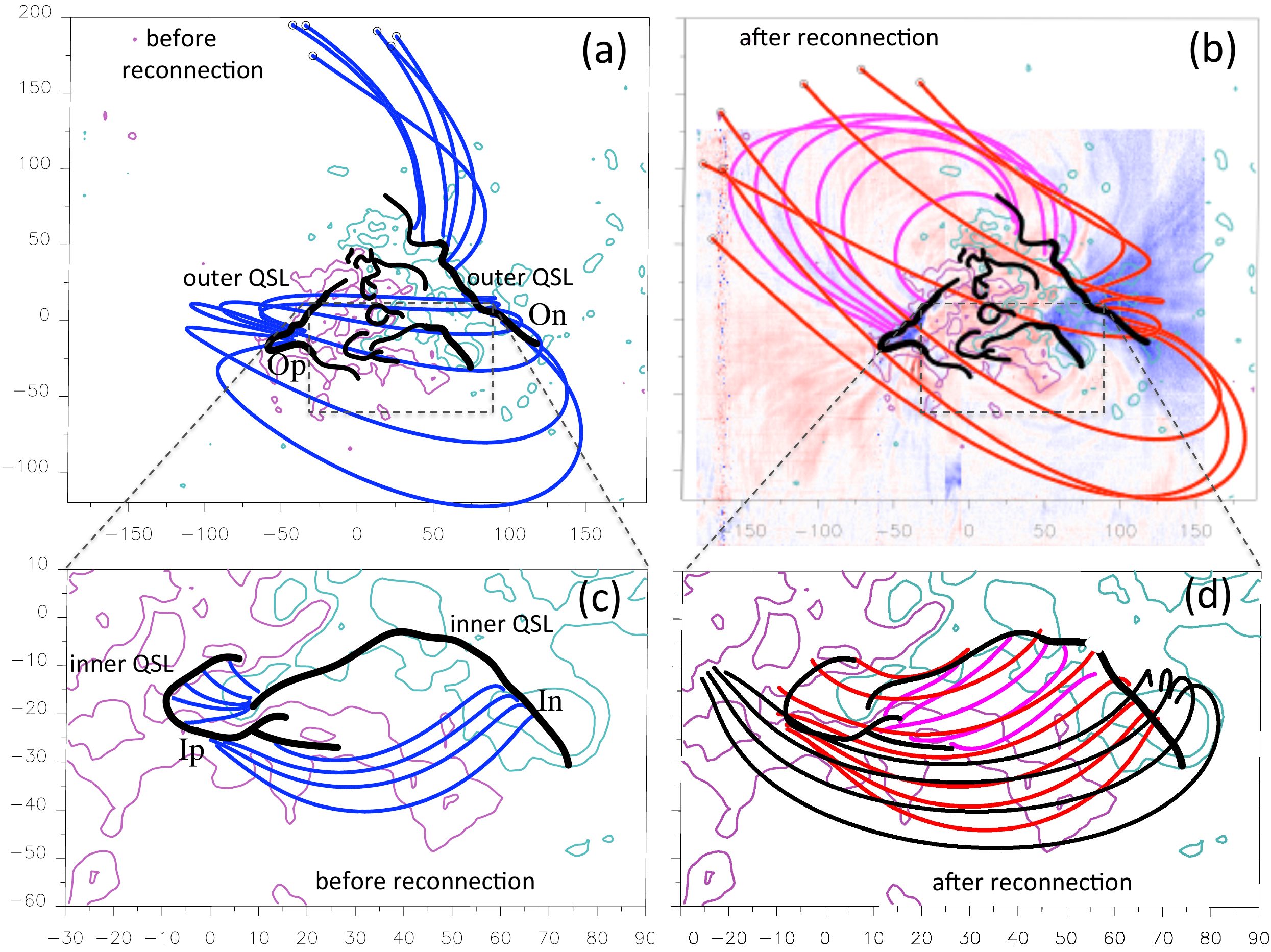}}
\caption{Magnetic field connectivity in the vicinity of the QSLs for the MDI magnetogram on 12 December 2007 at 3:26 UT. Panels (a) and (b) illustrate the pre-reconnection field lines in blue and the reconnected field lines in pink and red, respectively. These reconnected field lines would partially channel the EIS upflows located at the external eastern and western AR borders (Op and On; see also panel (d) in \fig{qsls-evol}). The corresponding EIS upflows are shown as background in panel (b).
The field lines that leave the drawn box correspond to large-scale loops connecting to positive quiet-Sun polarities (a circle is drawn at one of their ends). The dashed-line rectangle surrounds the region zoomed in panels (c) and (d). Panels (c) and (d) show the pre-reconnection and reconnected field lines using the same colors as in the top panels. These lines illustrate the reconnection process that would lead to the innermost EIS upflows. However, the red field lines in panel (d) do not follow the upflow spatial distribution; for this purpose, we have added a set of black lines in the vicinity of the western inner QSL (see \sect{qsls-detail} for an explanation). The convention for the magnetic field isocontours and axes is the same as in \fig{nulls}.
The spatial location of pre-reconnection and reconnected field-line footpoints on either side of the QSLs is typical of bipolar ARs \citep[see examples in][]{Demoulin97}.  }
\label{fig:reconnection}
\end{figure*}  

\subsection{Detailed Connectivity Analysis on 12 December}
\label{sect:qsls-detail}

In this section we present a scenario that, being consistent with the previously described observations and models, can explain why some sections of the QSL traces in AR 10978 are related to EIS upflows, while others are not and, also, why some upflow regions have different characteristics \citep[as shown by][]{Demoulin13}.
As an example, we show in detail the field-line connectivity at a particular date and time. As photospheric footpoint motions can provide forcing for the build-up of currents and induce reconnection along QSLs, we refer to the photospheric flows found in the AR (see \fig{flows-evol}), which facilitate the sequence of reconnections described below.

\fig{reconnection}a shows two sets of field lines starting on the outer QSL trace on the negative main AR polarity.  The northern field lines are anchored along the west side of the QSL trace and their opposite footpoints are located in quiet-Sun regions. These are large-scale loops probably filled with low-density plasma (low EUV emission). To the south of the same QSL trace we have a set of shorter higher-density loops within the AR. This is the situation we envision before reconnection occurs.

A large-scale photospheric flow pattern is present in the AR for several days, as shown for the period 10 to 12 December
(see \fig{flows-evol}b-d).  The main negative polarity is persistently moving to the south-west. As a result of this motion, the blue field lines anchored at the north of the outer QSL in \fig{reconnection}a may be forced to reconnect with the blue ones anchored to its south. After reconnection, we would have the two sets of field lines shown in \fig{reconnection}b. The process results, on one side, in the long red field lines with footpoints at the western border to the south of the outer QSL. The injection of high pressure plasma from the shorter pre-reconnected loops could drive the observed upflows at the western south border of the AR.
A similar process would drive, on the other side, the upflows on the eastern border of the main positive AR polarity along the reconnected pink field lines (\fig{reconnection}b).
We remark that we can identify both pre- and post-reconnection field lines in the same magnetic field extrapolation because we have computed both QSLs and the driving photospheric flows.  This is comparable to analyzing a snapshot in an MHD simulation knowing the velocity field direction. However, we have no way using these observations to identify a pair of pre-reconnection field lines evolving into a pair of post-reconnection field lines.

The process discussed above would happen provided a large enough asymmetry in plasma pressure exists between the pre-reconnected loops. In this way, and for as long as a forcing is at work, we would observe the persistent EIS upflows in AR 10978. We speculate that reconnection would occur across the QSLs in the slipping mode analyzed by \citet{Aulanier06} and further quantified by \citet{Janvier13}.

Comparison between upflow and QSL locations is not straightforward due to changing viewing angle and resulting projection effect, overlapping flows viewed in the optically thin corona, etc. However, if the reconnection process along QSLs lies at the origin of the EIS upflows, we expect that the projected spatial extension of the reconnected field lines at both AR borders compares to the spatial distribution of the observed Fe {\scshape xii} upflows. This is indeed the case for the pink reconnected field lines anchored in the main positive AR polarity (Op), but only partially so for the red ones anchored in the main negative polarity (In; \fig{reconnection}b).  However, field lines computed on the side and close to the QSL, so in lower $Q$ values such as the red field lines in \fig{qsls-evol}d, have a projected extension that compares well with the extension of the upflow region at the western AR border, as do the pink ones in the same panel in relation to the upflows at the eastern AR border. Both of these sets would correspond to previously reconnected lines at the outer QSLs if we take into account that the QSL trace will shift to the east (west) on the main negative (positive) polarity as reconnection proceeds.

Though the upflow region to the west of the AR in \fig{reconnection}b looks in projection as a single extended broad stream, it is in fact composed by at least two different streams (\fig{upflow-evol}). The easternmost border of this region lies in the vicinity of the inner QSL trace. In \fig{reconnection}c we show a set of blue field-lines that have been computed starting integration from the eastern side of the inner QSL trace on the negative polarity. Magnetic reconnection between lines in this set and those which are anchored on the western border of the inner QSL on the positive polarity, also drawn in blue, would result in the red and pink lines shown in \fig{reconnection}d. This reconnection process could be at the origin of the upflows located to the west of the inner QSL (at In). The projected extension of the reconnected red field-lines does not match the apparent extension of the upflows; in fact, for this set of lines we would expect upflows to the east of this QSL trace. As done for the outer QSL,
we integrated field lines anchored towards the west of the QSL trace. Such field lines, drawn in black in \fig{reconnection}d, have a  projected shape first directed to the west and then to the east. Plasma upflows along these lines would have the observed spatial distribution, i.e. we would observe stronger upflows close to the QSL trace fading towards its west. We also noticed that if we increased the absolute value of the $\alpha$ parameter in our LFFF model, keeping the same footpoints, the projected shape of the field lines was matching better the upflow extension. Then, we attribute the discrepancy between upflows and field-line projected shapes to the limitation of our LFFF model that does not represent well the probably higher-sheared small-scale AR magnetic field.

\begin{figure*} 
\centerline{\includegraphics[width=0.70\textwidth]{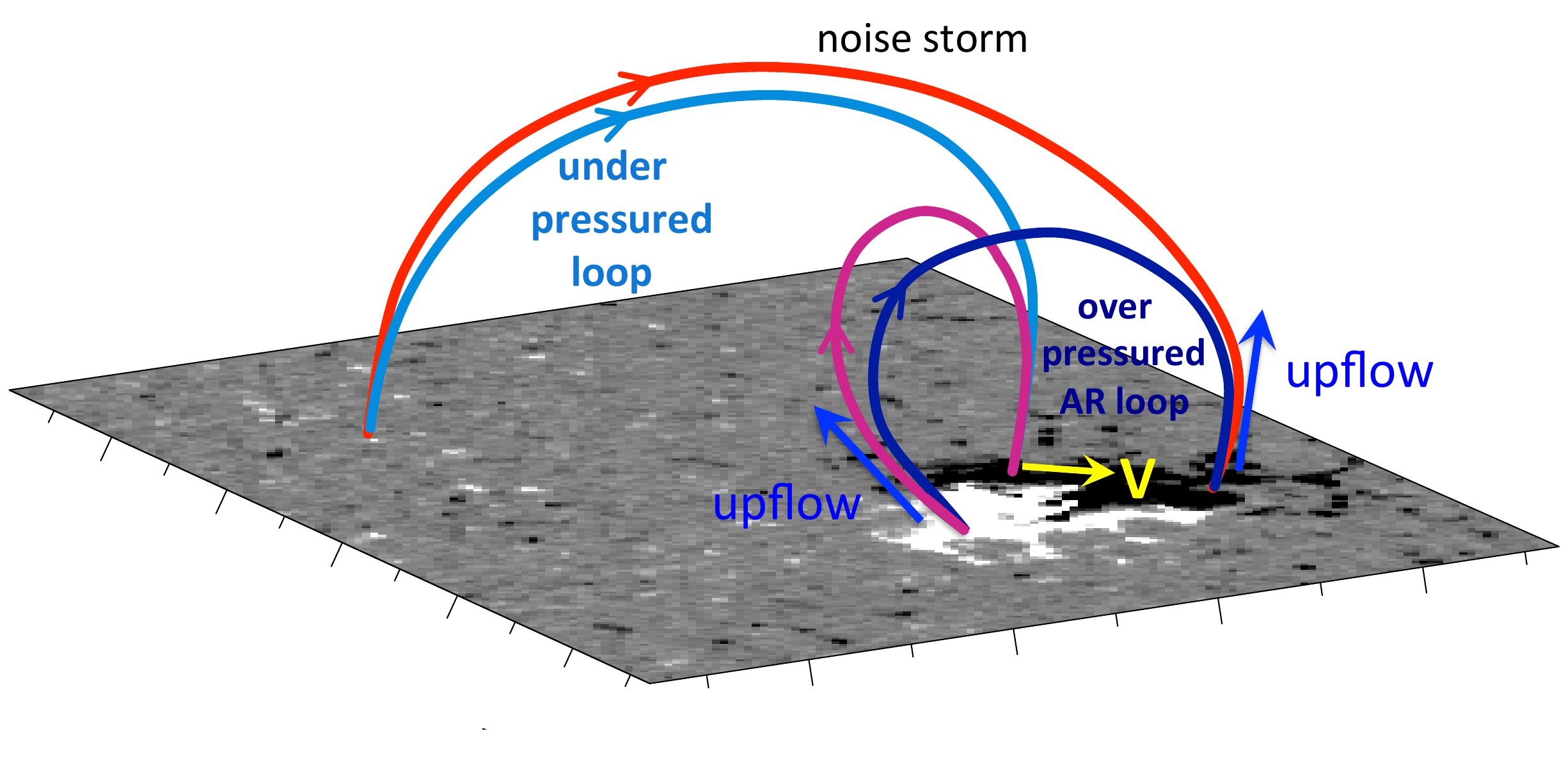}}
\caption{Perspective view of the AR and the magnetic field connectivity sketching the main processes involved. Four field lines,  corresponding to the four types of connectivity found in the coronal magnetic field model (\fig{reconnection}), are added on top of a magnetogram of the vertical field component.
The photospheric flows shown in \fig{flows-evol}, indicated by a yellow arrow, induce reconnection between over-pressured loops (dark-blue) located within the AR with under-pressured loops (light blue) connecting the AR to the quiet Sun by building up currents along QSLs in the corona. The pressure imbalance drives the upflows in the reconnected loops and the electrons accelerated during the quasi-steady sporadic reconnections generate a long-lived radio noise storm.
}
\label{fig:rec-summary}
\end{figure*}  

Our results explain the origin of all the upflow regions, shown in \fig{qsls-evol}, for 12 December 2007 as due to magnetic reconnection at QSLs. We have done similar detailed connectivity analyses for all the upflows shown in that figure. They seem to originate either by reconnection within the outer QSLs, which are associated to the large-scale magnetic field configuration of the AR, or in the inner QSLs that develop as new bipoles emerge and evolve during the AR disk transit.

 A two-step reconnection process was proposed by \citet{Mandrini14a} to explain the way AR 10978 upflowing plasma could access open field-lines and be observed by in situ instruments onboard satellites at L1. In that article we found evidence for the second step reconnection in the process, which opens the path for the closed-field confined plasma into the solar wind. Our present results offer proof of the first reconnection step, which is related to reconnection within the outer QSLs and provides a pathway for plasma originally confined along AR loops to flow into large-scale loops connecting to the quiet sun. The latter is summarized in \fig{rec-summary}.
Furthermore, our findings of a coherent evolution between magnetic field, QSLs, and upflows prove the concept first proposed by \citet{Baker09a} to explain the origin of EIS persistent blueshifts. In this article, we go one step forward and suggest how, due to the photospheric field evolution, only sections of the QSLs in the AR are linked to persistent plasma upflows.  Further evidence of a persistent reconnection process along QSLs in the AR can be revealed by radio observations of a persistent metric noise storm above AR 10978.

\begin{figure*} 
\centerline{\includegraphics[width=1.\textwidth]{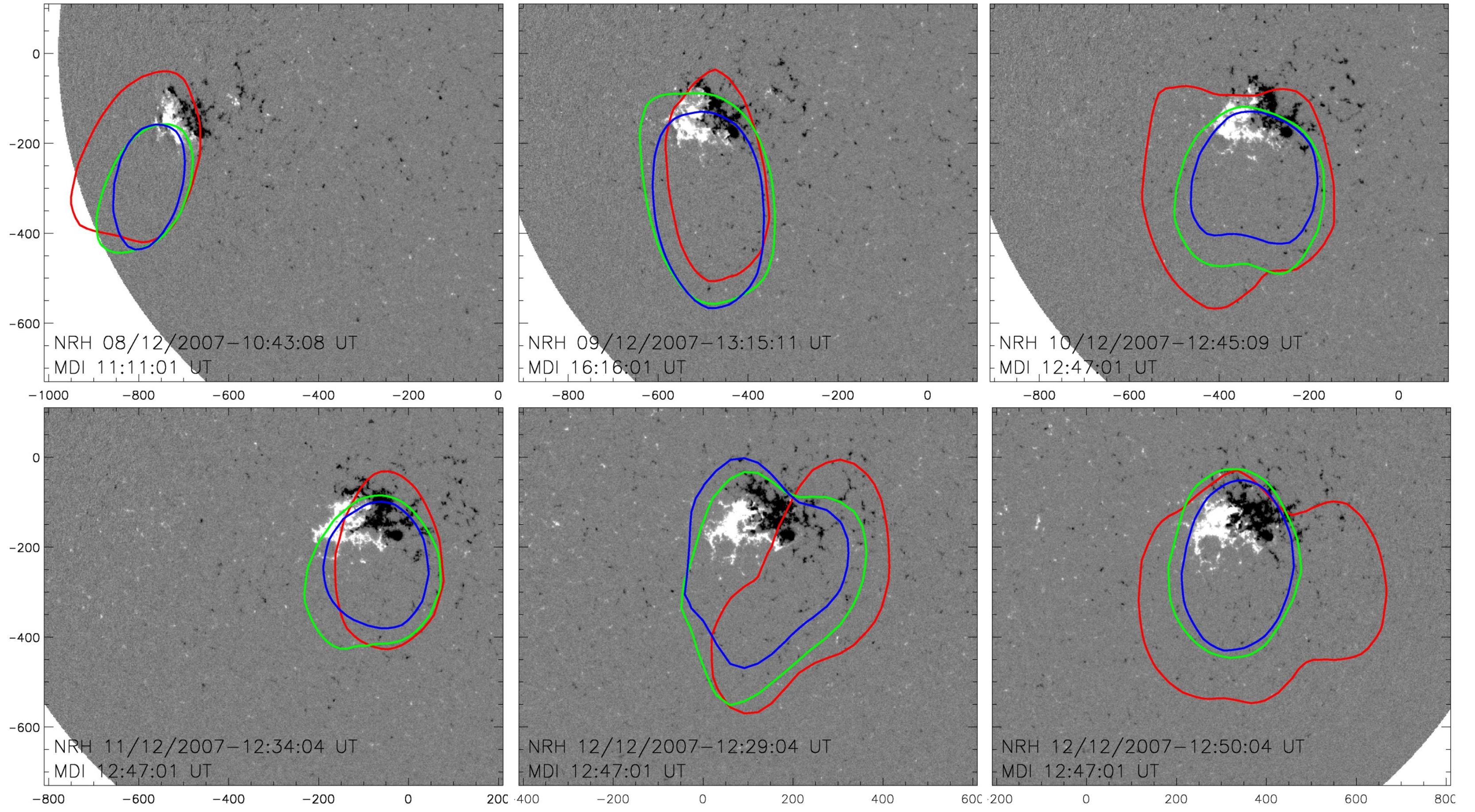}}
\caption{Radio noise storm sources over AR 10978 from  8 to 12 December 2007. The gray scale shows the magnetic field of the AR (MDI magnetograms).
Superimposed we show isocontours at 80\% of the maximum intensity (5 min integration)
of the radio sources at 432 MHz (blue), 408 MHz (green), and 327 MHz
(red). The selected radio observations show a typical distribution of the contours
during the full observing period.  The axes indicate the position on the Sun in arcsec.}
\label{fig:noise-evol}
\end{figure*}  

\section{Magnetic Field Topology and Noise Storms}
\label{sect:radio}

\subsection{Noise Storms: Characteristics}
\label{sect:noise-storms}

Following our study in \citet{Mandrini14a}, in this section we search for evidence of energy release at QSLs in wavelengths typical of radio noise-storms.
Noise-storms consist of a broadband continuum emission (bandwidth $\sim$200~MHz, around a central frequency of hundreds of MHz) lasting for hours to several days. Superimposed on this continuum, bursts of
short duration with much smaller bandwidth ($\sim$1~MHz) are observed.  The continuum component exhibits a slow intensity variability. \citet{Mercier15} analyzed radio observations in the range from 150 to 450 MHz with high spatial resolution and they concluded that noise storms show an internal structure with one or several compact cores embedded in a more extended and tenuous halo.
The emission is due to the presence of a suprathermal electron population (energies ranging from one to a few tens of
keVs) injected and trapped in extended coronal structures, i.e. noise storms require a mechanism that quasi-continously accelerates electrons in the solar corona.

The onsets of noise-storms or their enhancements are often related to changes in the overlying corona \citep{Kerdraon83} and to energy release in the underlying active region \citep{Raulin94,Crosby96}. These characteristics suggest that the plasma--magnetic field configuration is restructuring at the time and the place where the noise-storm is produced (see the next section).

\begin{figure} 
\centerline{\includegraphics[width=0.5\textwidth]{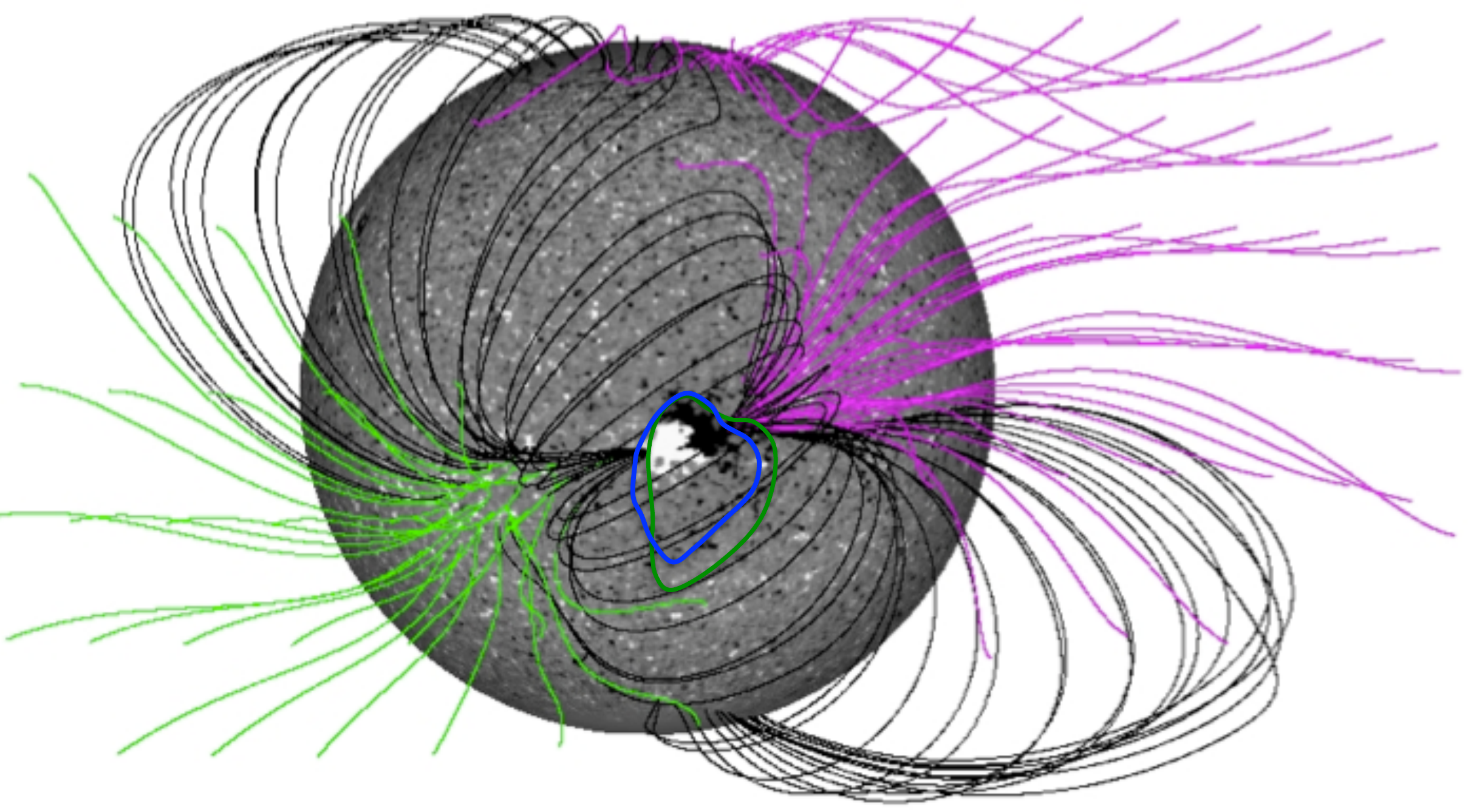}}
\caption{PFSS model of CR 2064 with AR 10978 located at the central solar meridian. The AR lies completely below the streamer belt and open field-lines from two coronal holes are seen to the south-east and north-west of the AR. The field-line color convention is such that black indicates closed lines and pink (green) corresponds to open lines anchored in the negative (positive) polarity field. The magnetic field is saturated above (below) 30 G (-30 G).
We have superimposed the NRH radio-emission isocontours at 432 MHz (blue) and 408 MHz (green) shown in \fig{noise-evol} on 12 December 2007.
The radio emission is concentrated over the closed field lines anchored at the border of the AR and encloses it.}
\label{fig:pfss}
\end{figure}  

\subsection{Noise-Storms: Evolution}
\label{sect:noise-storm-evol}

We analyze the radio emission registered by the Nan\c cay Radio Heliograph \citep[NHR,][]{Kerdraon97} during the transit of AR 10978 across the solar disk. At higher observation frequencies (327, 408, 432~MHz), the emission indicates the presence of  weak noise storms that remain almost unchanged during
the whole observing period. \fig{noise-evol} illustrates the evolution of the noise storm above the AR.
Radio emission contours at 80\% of the maximum intensity are displayed over the nearest in time MDI magnetogram; these contours were built using 5 min integrated data. They show no clear trend to strengthen (by increasing their sizes) or, conversely, to fade out (by diminishing their sizes).

At lower frequencies (150.9, 228~MHz), the radio emission contours embrace almost the full solar disk (not shown in \fig{noise-evol}) indicating that the radiation corresponds to background coronal emission of thermal origin.
Nan\c cay Decameter Array \citep[DAM, 10--80 MHz,][]{Lecacheux00} did not observe Type III bursts, which are known to be caused by  electrons that are accelerated outwards along the open coronal field-lines. DAM observes the upper corona from 0.7 to 3 $R_{\odot}$.  The lack of Type III emission seems to indicate that accelerated particles were not injected into open coronal structures during the observing periods as the AR crossed the disk.

\subsection{Noise Storms and Magnetic Field Connectivity}
\label{sect:noise-storm-topo}

The global coronal magnetic field of CR 2064 was modeled in the potential field source surface (PFSS) approximation (\fig{pfss}), assuming a current-free coronal field using as photospheric boundary condition a synoptic magnetogram. To close the upper boundary, PFSS models assume that the field becomes purely radial at a given height, called the source surface. This is a free parameter usually set to the value 2.5 $R_{\odot}$. The PFSS model in this article was computed with the Finite Difference Iterative Potential-Field Solver (FDIPS) code described by \citet{Toth11}, using the corresponding MDI synoptic magnetogram for CR 2064 as photospheric boundary condition.

As shown in \fig{pfss}, the noise-storm radio emission is concentrated over the closed AR field lines with an extension to the south due to projection effect. NRH isocontours fully enclose the AR. This spatial relation suggests that the same magnetic reconnection process that drives EIS upflows may accelerate the electrons that flow along the closed reconnected field lines originating the radio noise-storm emission.

\section{Summary and Conclusions}
\label{sect:conclusions}

Since the discovery of ubiquituous plamas upflows in EIS observations, several driving mechanisms were proposed. Among them were the impulsive heating at the footpoints of AR loops \citep{Hara08}, ``open'' magnetic funnels explaining coronal plasma circulation \citep{Marsch08}, chromospheric evaporation due to reconnection forced by flux emergence and/or braiding of field lines by photospheric motions \citep{DelZanna08}, expansion of large-scale reconnecting loops \citep{Harra08}, continual AR expansion \citep{Murray09}, and, more recently, reconnection between over-pressure AR loops and neighboring under-pressure loops \citep{Bradshaw11}.
 \citet{Baker09a} were the first to demonstrate the spatial relation between the location of upflows and QSLs at the border of a particular AR and to propose that magnetic reconnection at QSLs was at the origin of EIS upflowing plasma. \citet{vanDriel-Gesztelyi12} also found that EIS upflow regions at the border of an AR were cospatial with QSLs in another case study. However, an analysis of the spatial and temporal evolution of upflows and QSLs, which would provide strong support to the results found for individual examples, was still missing.

From an analysis of the evolution of the photospheric magnetic and velocity fields of AR 10978, as it transits the solar disk, combined with coronal magnetic field modeling and topology computations, we find that EIS upflow regions and QSLs evolve in parallel (\fig{qsls-evol}). Two sets of QSLs, called outer and inner (\figs{qsls-evol}{reconnection}), are found associated to EIS upflow regions with different characteristics \citep{Demoulin13}. All of the EIS upflows in AR 10978 seem to originate either by reconnection within the outer QSLs, which are associated to the large-scale magnetic configuration of the AR, or within the inner QSLs that develop as new bipoles emerge and evolve within the AR during its disk transit (see an example in \fig{reconnection}). The reconnection process in sections of the outer QSLs is forced by a large-scale photospheric flow pattern which is present in the AR for several days. In our proposed scenario, which is summarized in \fig{rec-summary}, upflows will be present provided a large enough asymmetry in plasma pressure exists between the pre-reconnected loops and for as long as a photospheric forcing is at work. A similar mechanism would be at work in sections of the inner QSLs, in this case forced by the emergence and evolution of the bipoles between the two main AR polarities. Furthermore, and within the limitations of our coronal field model, the projected extension of EIS upflows in AR 10978 match the projected shape of magnetic field-lines computed in the vicinity of QSLs.
Thus, our findings offer both observational and modeling support to the concept first put forward by \citet{Baker09a} and suggest how, due to the photospheric field evolution, only sections of the QSLs in the AR are linked to persistent plasma upflows.

Recent studies show that EIS upflowing plasma can gain access to open-field lines and be released into the slow solar wind via magnetic-interchange reconnection at magnetic null-points \citep[see e.g. ][]{vanDriel-Gesztelyi12}.
As shown in \fig{pfss}, AR 10978 is completely covered by closed streamer field-lines. Therefore, it seems unlikely that the upflowing plasma from AR 10978 can reach the solar wind; however, \citet{Culhane14} found signatures of plasma with AR composition at 1 AU
that apparently originated west of AR 10978. Based on a topological analysis of the global coronal magnetic field around AR 10978, \citet{Mandrini14a} proposed
a two-step reconnection process to explain the way a fraction of the AR upflowing plasma could access open field-lines.
Our present results describe the first reconnection step which occurs
within the outer QSLs and brings the plasma originally confined along AR loops to flow into large-scale loops connecting to the quiet Sun.
The second step reconnection in the process, modeled by \citet{Mandrini14a}, opens the path into the solar wind for the plasma confined into those large-scale loops.
We also find further evidence of this first step in radio observations. Comparison of the large-scale global coronal field (\fig{pfss}) to the location of a persistent metric noise-storm above AR 10978 observed by NRH, suggests that closed field lines reconnected within the outer QSLs may channel the accelerated electrons at the origin of the noise-storm.

A variety of magnetic configurations have been associated to EIS upflows located at AR borders.
Some upflows have been related to magnetic reconnection at magnetic null-points and associated separatrices \citep[\eg][]{DelZanna11}, in other cases no null points were present \citep[\eg][]{Baker09a} or reconnection at null points could not explain the majority of EIS upflows (this article), while there are other examples in which reconnection explaining upflows could happen at both null points and QSLs \citep[\eg][]{vanDriel-Gesztelyi12}. This is similar to what has been found for solar flares \citep[\eg][]{Demoulin94b,Mandrini10}. Furthermore, solar flares can be either confined or associated to coronal mass ejections (CMEs). In the later case, they have a direct impact on the interplanetary space \citep[\eg][]{Rouillard11b}.
The same happens with upflows that can remain confined within coronal loops or become outflows and access the solar wind
\citep[\eg][]{vanDriel-Gesztelyi12,Culhane14}.
On the other hand, solar flares are intrinsically impulsive events in which the magnetic energy, accumulated on the time scale of days, is released in a short time interval (typically, less than one hour) and the reconnection process is fast \citep{Shibata11}.
Conversely, upflows involve magnetic reconnection on the time scale of days. Upflows are driven by the long-term slow evolution of the AR magnetic field (emergence, large-scale velocity patterns, diffusion), i.e. upflows are not driven by a global instability of the magnetic field like flares and CMEs, but rather by a gradual evolution of the magnetic field.


\acknowledgements
C.H.M., G.D.C., F.A.N., and A.M.V. acknowledge financial support from the Argentine grants PICT 2012-0973 (ANPCyT) and PIP 2012-0403 (CONICET). C.H.M. and G.D.C. thank grant UBACYT 2013-0100321 (UBA).
The research leading to these results has received funding from the European Commission's Seventh Framework Programme under the grant agreement No. 284461 (eHEROES project). L.v.D.G. and D.B. acknowledge support by STFC Consolidated Grant ST/H00260/1.
L.v.D.G.'s work was supported by the Hungarian Research grant OTKA K 109276.
This work was partially supported by one-month invitation to P.D. to visit the Instituto de Astronom\'\i a y F\'\i sica del Espacio and by  one-month invitation to C.H.M. to visit the Paris Observatory.
C.H.M., G.D.C., and A.M.V.  are members of the Carrera del Investigador Cient\'\i fico (CONICET).
F.A.N. is a fellow of CONICET.

\bibliographystyle{apj}
\bibliography{dec-rotation-qsls}
\end{document}